\documentclass[aip,pof,amsmath,amssymb,reprint]{revtex4-1} \synctex=1 

\usepackage{cmap}
\usepackage[OT1]{fontenc}
\usepackage{graphicx}
\usepackage[colorlinks=true,linkcolor=blue]{hyperref}
\usepackage{hyperref}
\usepackage{bm}

\begin{document}


\title{Fingering instability of a suspension film spreading on a
  spinning disk} \author{Mayuresh Kulkarni}

\author{Subhadarshinee Sahoo}

\author{Pankaj Doshi} \altaffiliation{Current address: Pfizer, Inc.,
  Groton, Connecticut 06340, USA.} \email{pankaj.doshi@pfizer.com}
\author{Ashish V. Orpe} \email{av.orpe@ncl.res.in}

\affiliation{Chemical Engineering Division, CSIR-National Chemical
  Laboratory, Pune 411008, India}
  

\begin{abstract} The spreading of a thin film of suspension on a
  spinning disk and the accompanying contact line instability is
  studied through flow visualization experiments. The critical radius
  for the onset of instability shows an increase with
  increase in the particle fraction ($\phi_{p}$) before decreasing
  slightly at the highest value of $\phi_{p}$ studied, while the instability
  wavelength ($\lambda$) exhibits a non-monotonic dependence. The
  value of $\lambda$ is close to that for a partially wetting liquid
  at lower $\phi_{p}$, it decreases with increasing $\phi_{p}$ to a
  minimum before increasing again at largest $\phi_{p}$. The
  non-monotonic trends observed for $\lambda$ are discussed in
  light of the linear stability analysis of thin film equations
  derived for suspensions by Cook {\it et al.} [“Linear stability of particle-laden
  thin films,” Eur. Phys. J.: Spec. Top. {\bf 166}, 77 (2009)] and Balmforth {\it et al.}  		
  [“Surface tension driven fingering of a viscoplastic film,” J. Non Newtonian
  Fluid Mech. {\bf 142}, 143 (2007)]

\end{abstract}

\maketitle

\section{Introduction}
\label{sec:introduction}

The spreading of a thin, viscous, film of liquid under external
forcing (gravity or centrifugal) comprises of two regions: a flat
region, the dynamics of which are determined by the balance between
driving force and viscous dissipation and a second region near the
advancing front, the shape of which is governed by the interplay
between the liquid surface tension and the driving force. This second
region, termed capillary ridge, is susceptible to perturbations in the
transverse direction eventually growing as
fingers.~\citep{Emslie1958,Huppert1982,Troian1989,Melo1989,Brenner1993,Fraysse1994,Spaid1996,Schwartz2004} The presence of particles in such a spreading thin film of liquid (i.e., a
suspension or a slurry) is encountered in different
applications~\citep{Mintova2001,Mackaplow2006,Pichumani2013} and can
exhibit several complexities during spreading and ensuing fingering
patterns.~\citep{Bruyn2002,Zhou2005,Holloway2010} 
 
For suspensions which exhibit yield stress behavior, the capillary
ridge gets stabilized by the innate yield strength of the fluid
allowing the film to spread to a larger area before the instability
ensues.~\citep{Bruyn2002,Balmforth2007} Linear stability analysis of
the thin film equations for such yield stress fluids, incorporating a
suitable stress constitutive equation, correctly predicts the observed
behavior.~\citep{Balmforth2007} In certain cases, however, the
instability can be simply due to local yielding of the
suspension~\citep{Holloway2010} and not the usual contact line
instability along the advancing film front. An enhancement in the
fingering instability, compared to the base suspending liquid, is
predicted using linear stability analysis of thin film equations for
hard sphere, non-Brownian, well-mixed
suspensions.~\citep{Cook2009} This enhancement is found to be slightly
subdued for the case of settling, heavier particles, also observed in
experiments.~\citep{Zhou2005} 
 
Here, we report interesting and novel experimental observations on a
thin film of suspension spreading on a horizontal spinning disk under
the influence of centrifugal forcing. The suspending liquid is
partially wetting and is not easily amenable to spreading and
instability, owing to its surface minimizing tendency. However, the
suspension of neutrally buoyant hard spheres in this liquid, at
increasing particle fractions ($\phi_{p}$), exhibits greater tendency
to spread before the initiation of instability compared to the
spreading of the suspending liquid. Further, the wavelength of contact
line instability exhibits a non-monotonic behavior within the same
range of $\phi_{p}$. The wavelength is large for $\phi_{p} \le 0.4$,
it decreases to a minimum in the range $0.4 < \phi_{p} < 0.5$ before
increasing again at higher $\phi_{p}$. The behavior at the lower and
highest values of $\phi_{p}$ seems to mimic, respectively, that of a
partially wetting liquid and a yield stress suspension, while the
behavior at intermediate values of $\phi_{p}$ seems to mimic that of a
highly wetting liquid. We explain the observed phenomena based on the
predictions of linear stability analysis of the thin film equations
for suspensions presented previously.~\citep{Balmforth2007,Cook2009}

\section{Experimental details}
\label{sec:experimental-details}

\begin{figure} \includegraphics[scale=0.45]{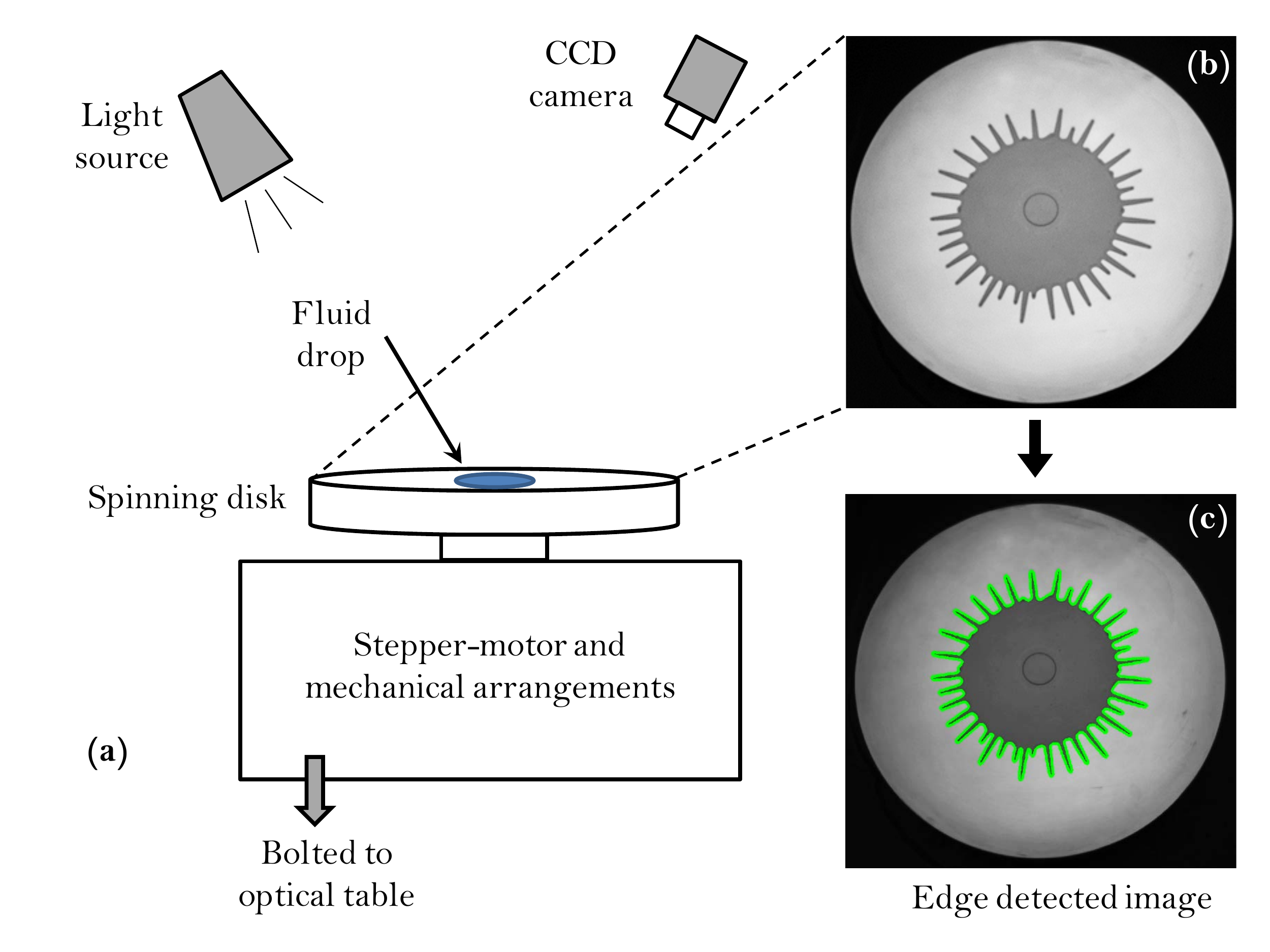}
  \caption{\label{fig:schematic}Schematic of the experimental set
    up. (a) Spinning disk assembly and image acquisition system (b)
    Sample image (for PDMS) taken at certain stage of spreading (c) Same
    image as in (b) but superimposed with the edge detected
    coordinates.}
\end{figure}

The experimental assembly shown in Fig.~\ref{fig:schematic} consists
of a flat, aluminium disk of diameter $15$ cm driven by a computer
controlled DC stepper motor. The disk can be rotated in the range
$50-10000$ revolutions per minute (rpm). The desired speed is achieved
from rest using an acceleration of 1000 rpm/s. The feedback
controller mechanism ensures that the random fluctuations in the
rotational speed are within $2$\% for $250$ rpm while they are within
$5$\% for $1000$ rpm.

Three different fluids, two Newtonian liquids and a suspension, were
used in the experiments, the properties of which are given in
Table~\ref{tab:table1}. The suspension was prepared by immersing glass
beads (of an approximate normal size distribution with mean diameter
$d = 52 \mu$m  and standard deviation $11 \mu$m, and density $2.45$ g/cc as
procured from Potters, Inc.) in a liquid of matching density prepared
from a mixture of LST liquid (density $2.85$ g/cc) and glycerol
(density $1.26$ g/cc). The density matching between particles and
liquid is attained to the accuracy of three decimal places. LST
(Lithium heteropolytungstates) is a water soluble heavy mineral
liquid, typically used for particle separations.  Suspensions of
varying $\phi_{p} $ ($0.1-0.55$) were prepared by mixing different
amounts of glass beads in the glycerol-LST mixture. The viscosities of
all the fluids, including suspensions, were obtained using the steady
state rheology measurements carried out in a stress controlled
rheometer. For all values of $\phi_{p}$ studied, the suspensions
exhibit a Newtonian behavior for shear rates upto $10^{3} $ s$^{-1}$
(which are much more than those encountered in experiments) and the
measured values of viscosity obey the well known Krieger-Dougherty
equation for hard sphere suspensions. To measure the static contact
angle, a drop of fluid was placed on the horizontal substrate. The drop
was imaged by a camera placed sideways and in the plane of the
substrate. The image was analyzed using
ImageJ software to obtain the 
static contact angle. The values and corresponding errors reported in
Table~\ref{tab:table1} represent average over $20$ independent
measurements. 

\begin{table} 
  \begin{ruledtabular} 
    \begin{tabular}{ccccc} 
      Fluids & Density & Viscosity & Volume & Static contact angle \\ 
      & (g cm$^{-3}$) & (Pa s) & (cm$^{3}$) & (deg)\\ \hline 
      Glycerol & 1.26 & 0.896 & 1.1 & $ 68\pm3$ \\ 
      PDMS & 0.91 & 0.173 & 0.8 & $7\pm3$ \\ 
      LST+Glycerol ($\phi_{p} = 0.0$) & 2.45 & 0.022 & 1.1 & $65\pm3$ \\ 
      Suspension ($\phi_{p}=0.100$) & 2.45 & 0.027 & 1.1 & $65\pm3$ \\ 
      Suspension ($\phi_{p}=0.200$) & 2.45 & 0.043 & 1.1 & $65\pm3$ \\ 
      Suspension ($\phi_{p}=0.300$) & 2.45 & 0.064 & 1.1 & $65\pm3$ \\ 
      Suspension ($\phi_{p}=0.400$) & 2.45 & 0.096 & 1.1 & $65\pm3$ \\ 
      Suspension ($\phi_{p}=0.425$) & 2.45 & 0.132 & 1.1 & $65\pm3$ \\ 
      Suspension ($\phi_{p}=0.450$) & 2.45 & 0.162 & 1.1 & $65\pm3$ \\ 
      Suspension ($\phi_{p}=0.475$) & 2.45 & 0.220 & 1.1 & $65\pm3$ \\ 
      Suspension ($\phi_{p}=0.500$) & 2.45 & 0.295 & 1.1 & $65\pm3$ \\ 
      Suspension ($\phi_{p}=0.525$) & 2.45 & 0.350 & 1.1 & $65\pm3$ \\ 
      Suspension ($\phi_{p}=0.550$) & 2.45 & 0.763 & 1.1 & $65\pm3$ \\ 
    \end{tabular} 
  \end{ruledtabular}
  \caption{\label{tab:table1}Physical properties of the fluids used in
    the experiments. The surface tensions of PDMS, Glycerol and 
    suspending liquid (LST$+$glycerol) measured using pendant drop
    method are, respectively, $19.2$ dyn cm$^{-1}$, $64.2$ dyn
    cm$^{-1}$ and $70$ dyn cm$^{-1}$. It is not possible to measure
    the surface tension of suspensions quite accurately.}
\end{table} 

In every experiment, a small drop of Polydimethylsiloxane (PDMS)
($0.8$ ml) or glycerol ($1.1$ ml) or suspension ($1.1$ ml) of
particular value of $\phi_{p}$ was placed at the center of the
disk. The disk surface, before every experiment, was washed multiple
times using soap solution and rinsed with DI water followed by acetone
to remove any traces of fluid used from the previous experiments. It
was then mounted exactly horizontally on the spinning assembly. This
protocol ensured the reproducibility of the experimental
results. Experiments were carried out for four different rotational
speeds of the disk: $250$, $500$, $750$, and $1000$ rpm. Each
experiment was repeated $5$ times and the results presented are
averages over these experiments.

The surface of the disk is illuminated from above using a bright
halogen lamp. The motion of fluid on the surface of the aluminium disk
is captured from above using a high speed camera with an exposure of
150 $\mu$s to acquire sharp images. A small amount of fluorescent dye
was added to the transparent fluid for ease in the fluid visualization
and image analysis using better contrast (see inset of
Fig.~\ref{fig:schematic}(b) showing a fluid, appearing in dark color,
which has spread to a certain extent). The addition of the fluorescent
dye changes the surface tensions of PDMS, glycerol, and LST-glycerol
mixture by $0.05$\%, $0.3$\%, and $3.3$\%, respectively, which is small
enough to induce any qualitative changes in the observed behavior. The
acquired images were analyzed to obtain the edges of the spreading
fluid using standard edge detection algorithms. The edges were
detected to an accuracy of $\pm 0.15$ mm (see
Fig.~\ref{fig:schematic}(c)). The edge detection data were used to
calculate several quantities, viz., effective radius at different radial
locations (i.e., different times), spreading rates, instability
wavelengths, number of fingers.

\section{Results and discussion}
\label{sec:results-discussion}

\subsection{Spreading behavior}
\label{sec:spreading-behavior}

\begin{figure} \includegraphics[scale=0.55]{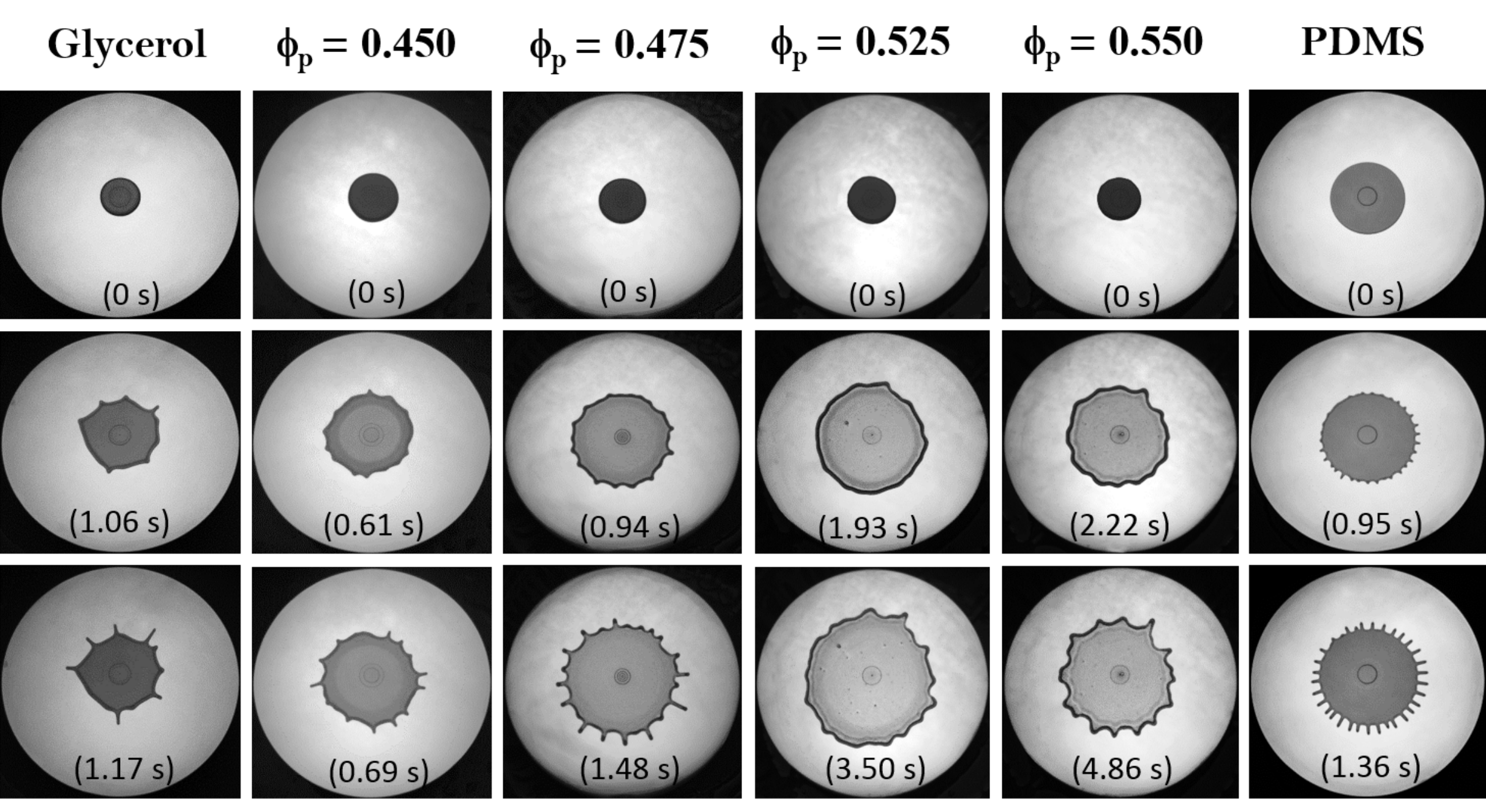}
  \caption{\label{fig:spreading_500}Static images for a few fluids
    obtained at different times (or different radial positions) for a
    rotational speed of $500$ rpm. Each column corresponds to a
    particular fluid while each row corresponds to the state of the
    fluid at a particular time (embedded in each image). The
    suspension images are flanked in the middle by images for
    partially wetting glycerol as first column and that for the
    completely wetting PDMS as the last column. Initial volume is
    $1.1$ ml for all fluids except PDMS for which $0.8$ ml was used.}
\end{figure}

Figure~\ref{fig:spreading_500} shows the spreading of partially
wetting glycerol, completely wetting PDMS and suspensions ($\phi_{p} =
0.45, 0.475, 0.525, 0.55$) at a rotational speed of $500$ rpm. Each
column corresponds to the spreading of a particular fluid at different
times. Each row corresponds to the same degree of deformation of the
drop boundary for all the fluids. The degree of deformation is defined
as $[(R_{1}-R_{2})/R_{2}] \times 100$, where $R_{1}$ is the largest
radius for the drop (distance of the farthest point from the axis of
rotation) and $R_{2}$ is the radius of a circle centered on the axis
of rotation and having same area as that covered by the drop. The rows
$2$ and $3$, thus, correspond to a deformation of $10$\% and $20$\%,
respectively, for all fluids.

The drop of PDMS, when placed on the disk, occupies a larger area
compared to the drop of glycerol and suspensions even if the volume
used is smaller (row $1$ in Fig.~\ref{fig:spreading_500}). This is due
to high wettability of PDMS (small contact angle) with respect to the
disk surface, which spreads to occupy a larger area. The suspension
drop (for all $\phi_{p}$) occupies nearly the same area of the disk as
the glycerol drop. This suggests that the suspending liquid, which has
the same wettability as glycerol (nearly same contact angles as shown
in Table~\ref{tab:table1}) primarily determines the final static
configuration and the particles have relatively lesser influence. The
finer details of the approach to this final static
state,~\citep{Han2012,Jeong2010} and the possible particle influence
therein, are not within the scope of this work and hence was not
studied.

Once the rotation of the disk is started ($500$ rpm as shown in
Fig.~\ref{fig:spreading_500}), the drops of glycerol (first column)
and PDMS (sixth column) start to spread leading towards the formation
of instability along the circumference (second row) and its further
growth into fingers at later times (3rd row). The radius of the base
circular region of the drop at which instability first appears, does
not seem to increase significantly with the fluid continuously flowing
outward through the fingers which eventually reach the periphery. This
overall behavior, inclusive of the instability wavelength (or number
of fingers formed), is very much in accordance with the previous
studies for Newtonian liquids.~\citep{Fraysse1994,Melo1989}
 
The suspension, for all values of $\phi_{p}$, spreads to a larger area
before the instability is initiated compared to the spreading of a
partially wetting liquid when the disk is rotated. Note that the
suspending liquid is partially wetting, like glycerol (see relevant
details in Table~\ref{tab:table1}). The extent of spreading, before
the initiation of instability, increases gradually up to
$\phi_{p}=0.525$ and then decreases slightly at the highest particle
fraction ($\phi_{p}=0.55$). This indicates
stabilization of the contact line due to the increasing presence of
particles which is qualitatively similar to that observed during the
spreading of a viscoelastic fluid~\citep{Spaid1997} and clay
suspensions which exhibit a finite yield
stress.~\citep{Bruyn2002,Balmforth2007} For the values of $\phi_{p}$
studied over here, the suspensions, however, do not exhibit a yield
stress or show a viscoelastic behavior as ascertained from the bulk
rheology measurements. Further, when compared to pure liquids, the
base circular region of suspension (for any particular $\phi_{p}$)
continues to spread significantly post the instability
initiation. Increasing the rotational speed beyond $500$ rpm does not
qualitatively change the spreading behavior, however, lowering the
rotational speed reduces the spreading tendency significantly. The
qualitative behavior of the fingering instability;
however, remains the same for different rotational speeds as shown
later.

\begin{figure} \includegraphics[scale=0.5]{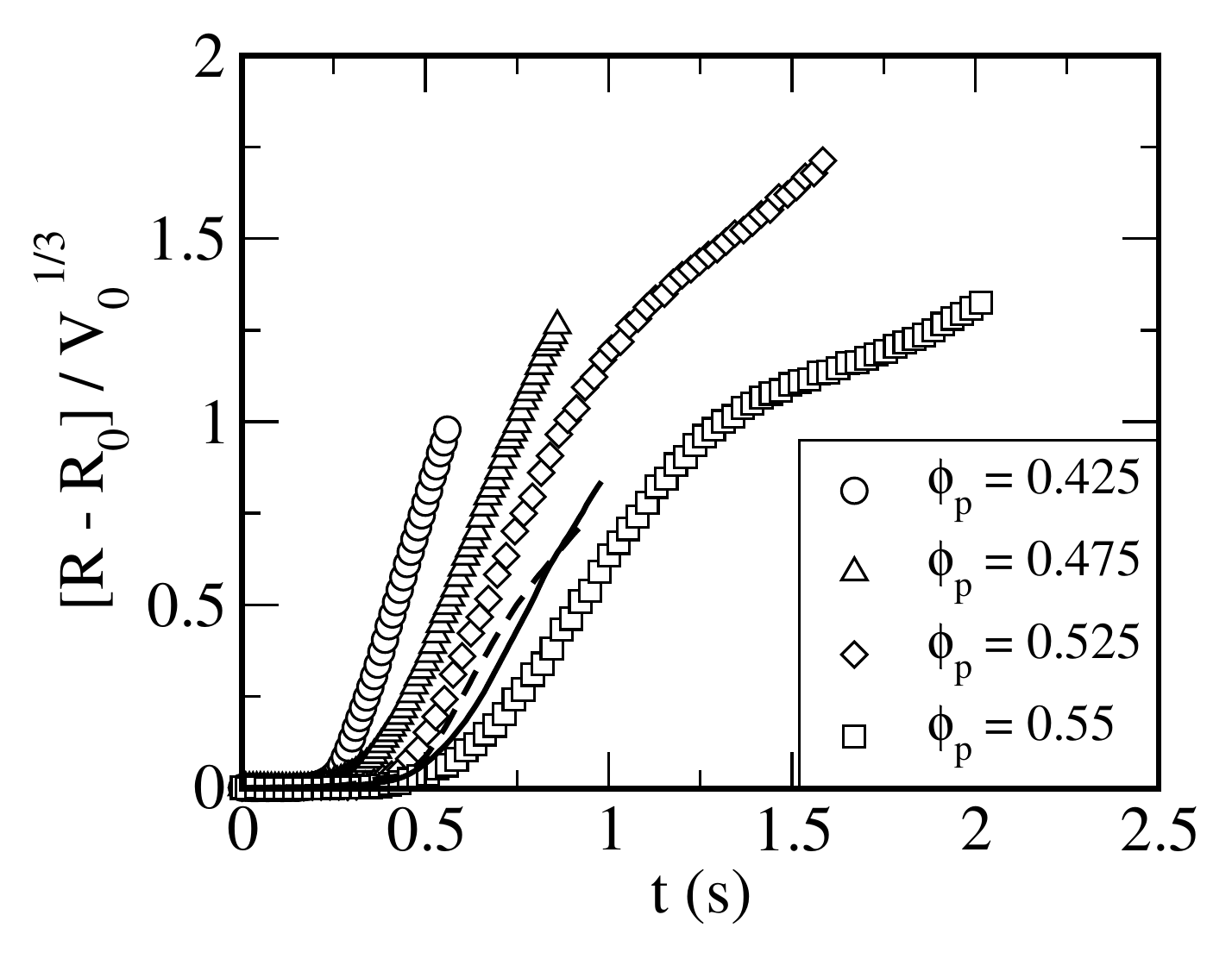}
  \caption{\label{fig:spreadrate}Spreading of a drop from
    its initial position for $500$ rpm. Solid line denotes the
    spreading of glycerol while the dashed line denotes the
    spreading of PDMS. The profiles represent the drop evolution
    from its initial position (row 1 in Fig.~\ref{fig:spreading_500})
    upto the critical radius (row 2 in Fig.~\ref{fig:spreading_500}).}
\end{figure}

The time evolution of the suspension drop from its initial state up to
the critical radius is shown in Fig.~\ref{fig:spreadrate} for the
rotational speed of $500$ rpm along with the evolution of the drop of
PDMS and glycerol. The results for the spreading/growth post-instability 
are shown later. The extent of spreading ($R-R_{0}$)
is normalized by the cube root of initial volume ($V_{0}$). Here, $R$
is the radius of the spreading drop and $R_{0}$ is the radius of the
drop in its initial state.  The final point in each profile
corresponds to critical radius (i.e., images in the second row shown in
Fig.~\ref{fig:spreading_500}). The rate of spreading of the suspension
drop decreases with increase in the particle fraction which can be
expected~\cite{Ward2009} given the increase in the viscosity (see
Table~\ref{tab:table1}). For the highest particle fraction
($\phi_{p}=0.55$) studied, the spreading rate is even lower than that
observed for the two liquids. The extent of spreading ($R-R_{0}$) for
the suspension drop, however, increases continuously with $\phi_{p}$
before decreasing slightly at $\phi_{p}=0.55$. Further, the time to
attain the critical radius increases monotonically with increase in
$\phi_{p}$. The overall behavior seems to arise due to combined effect
of viscosity as well as the ability of a fluid to spread. For
instance, the viscosities of PDMS and suspension ($\phi_{p} = 0.45$)
are nearly the same and both possess very different static contact
angles, but the extent of spreading for same degree of deformation
seems to be similar. Similarly, the viscosities of glycerol and
suspension ($\phi_{p} = 0.55$) are nearly the same and both are partially
wetting (large static contact angles), but still the suspension
spreads to a larger extent. Finally, the spreading rates and extent of
spreading for PDMS and glycerol, which have quite different
viscosities and wettabilities, are nearly the same.

\subsection{Instability and finger growth}
\label{sec:inst-fing-growth}

We, now, discuss the characteristics of the contact line instability
accompanying the spreading of the film. This instability ensues due to
a capillary ridge region formed near the advancing front of the film,
which is susceptible to perturbations in transverse direction
eventually growing as fingers associated with a prescribed
wavelength.~\citep{Huppert1982,Troian1989,Melo1989,Brenner1993,Fraysse1994}
The radius of the spreading film at which the instability first ensues
is defined as critical radius ($R_{c}$). We measure $R_{c}$
corresponding to the $10$\% deformation of the contact
line.~\citep{Fraysse1994} The value is normalized using the cube root
of the initial drop volume ($V_{0}$) which is not the same for all
fluids.

\begin{figure} \includegraphics[scale=0.5]{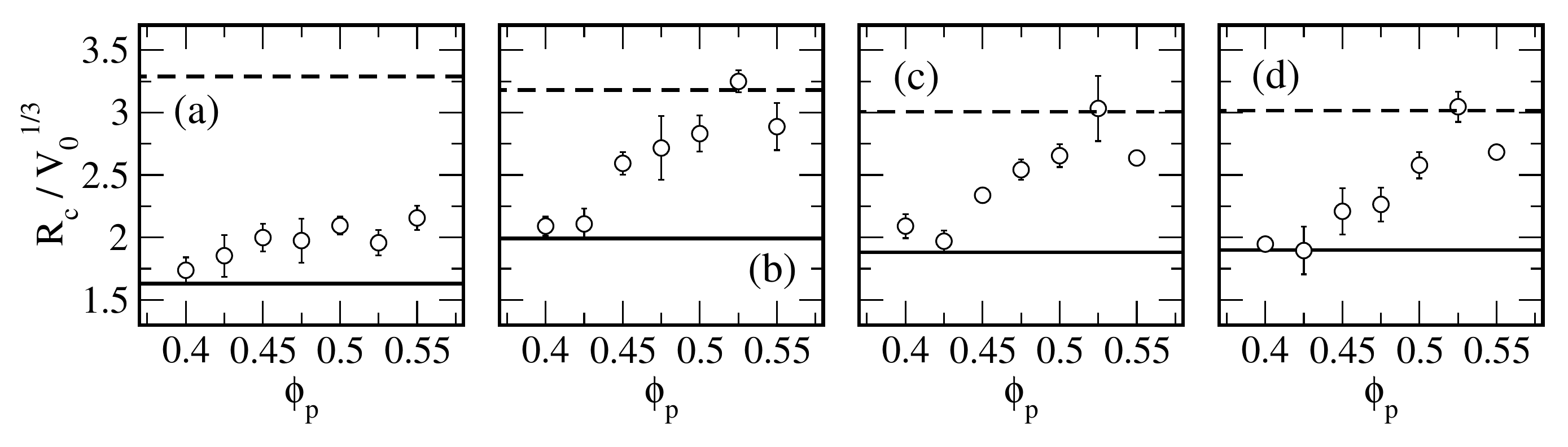}
  \caption{\label{fig:Rc}Normalized critical radius ($R_{c}$) at (a)
    $250$ rpm, (b) $500$ rpm, (c) $750$ rpm and (d) $1000$ rpm.  Open
    circles denote the data for suspension. Error bars denote
    deviations obtained by averaging over five independent data
    sets. Values of $R_{c}$ for glycerol and PDMS are shown,
    respectively, as solid and dashed lines.}
\end{figure}

Figure~\ref{fig:Rc} shows the variation of the normalized critical
radius with $\phi_{p}$ for all the rotational speeds studied. The
critical radius for PDMS decreases slightly over the range of
rotational speeds employed, which is in accordance with previously
observed behavior.~\cite{Fraysse1994} For all other fluids,
the critical radius increases with
increase in rotational speed from $250$ rpm to $500$ rpm, beyond which
it remains nearly constant. Further, the critical radius for glycerol
(shown as a solid line) is much lower than PDMS (shown as a dashed
line) for all the rotational speeds which is expected given that PDMS
is far more wetting than glycerol and hence can spread to a larger area
before developing instabilities (compare first and sixth column in
Fig.~\ref{fig:spreading_500}). The value of $R_{c}$ at
$\phi_{p} = 0.4$ and lower (not shown) is quite close to that for the
partially wetting glycerol for all the rotational speeds which
suggests that the presence of particles has a negligible influence on
the instability behavior for these values of $\phi_{p}$. For these
cases, the instability is, then, initiated due to the usual contact
line instability of the partially wetting suspending liquid. Beyond
$\phi_{p} = 0.4$, the value of $R_{c}$ increases steadily (except at
$\phi_{p} = 0.425$ for some rotational speeds which is not clear)
towards that for PDMS, with a slight decrease at the highest $\phi_{p}$
for rotational speeds of $500$ rpm and above. The rate of increase
is smaller for $250$ rpm compared to that for higher rotational speeds.

\begin{figure} \includegraphics[scale=0.5]{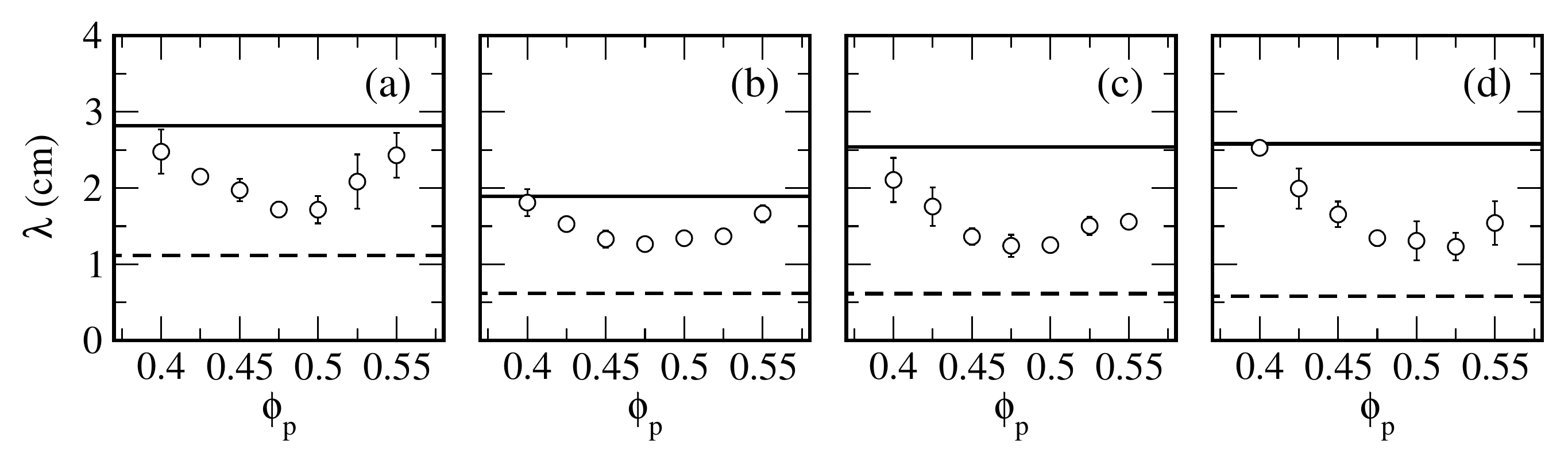}
  \caption{\label{fig:lam}Wavelength of the instability ($\lambda$)
    for (a) $250$ rpm, (b) $500$ rpm, (c) $750$ rpm and (d) $1000$
    rpm. Open circles denote the data for suspension. Error bars
    denote deviations obtained by averaging over five independent data
    sets. Values of $\lambda$ for glycerol and PDMS are shown,
    respectively, as solid and dashed lines.}
\end{figure}

The wavelength ($\lambda$) of the instability is related to the
experimentally determined critical radius ($R_{c}$) and number of
ensuing fingers ($N_{f}$) as counted from the images, by the geometric
relation~\citep{Fraysse1994} $\lambda = (2 \pi R_{c})/N_{f}$. In
Fig.~\ref{fig:lam}, we show the wavelength obtained by this relation
for different fluids using the experimentally determined values for
$R_{c}$ and the number of fingers (counted from the images) for all
the four rotational speeds employed. All the measurements are carried
out at a radial location where the degree of deformation is
$10$\%. The secondary fingers which form at later times (larger degree
of deformation or spreading) between the existing fingers are not
included over here. PDMS has lower surface tension and breaks into
several fingers associated with smaller instability
wavelength. Glycerol, on the other hand, has higher surface tension
and breaks into lesser fingers associated with larger instability
wavelength to minimize the surface energy. This behavior is very well
known in the literature and the values of the wavelength for both the
liquids are in accordance with the predictions obtained through the
linear stability analysis of thin film
equations.~\citep{Huppert1982,Melo1989,Troian1989,Fraysse1994}

The data for suspension exhibit a very interesting non-monotonic
dependence on increasing $\phi_{p}$ for all the rotational speeds. The
decrease in the wavelength is in qualitative agreement with the linear
stability analysis of thin film equations by Cook et
al.\citep{Cook2009} These authors developed a continuum two-phase
model for suspension, including concentration gradient driven
diffusion and shear-induced migration of particles.  This model is
able to capture the decrease in the instability wavelength for a
suspension film compared to that for the film of the suspending
liquid. In our experiments, this raises the possibility of the
presence of particle concentration gradients within the capillary
ridge which can cause diffusion of particles and consequently lower
wavelengths. The primary reason behind the formation of such a
concentration gradient in the system, if at all present, is however
not clear and needs to be investigated further. The higher wavelengths
observed in Fig.~\ref{fig:lam} at higher values of $\phi_{p}$ seem to
be in qualitative agreement with the predictions of linear stability
analysis of thin film equations for yield stress suspension by
Balmforth et al.\citep{Balmforth2007} The model, incorporating a
Bingham stress constitutive equation, captures the increasing
wavelength for a suspension film compared to the film of the
suspending liquid, through the increasing yield stress. For the higher
concentrations studied in our experiments, this raises the possibility
of particle crowding within the capillary ridge, which will ensue a
contact network and impart a yield stress, leading to the observed
higher wavelengths. Again, the exact mechanism behind the crowding of
the particles in the ridge, if at all present, needs further
investigation.

\begin{figure} \includegraphics[scale=0.5]{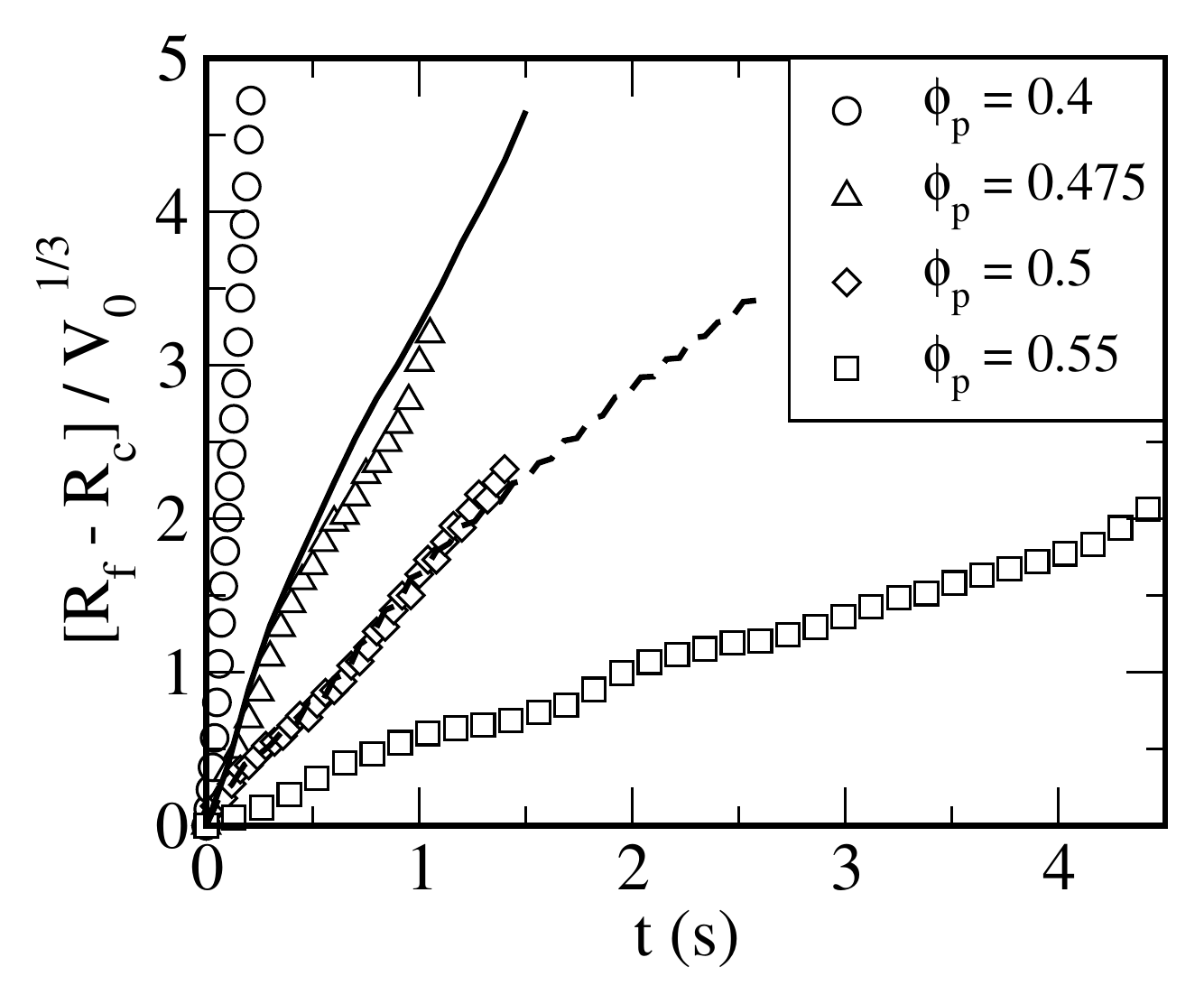}
  \caption{\label{fig:front_position}Normalized growth of the fastest
    moving finger for $500$ rpm. Solid line denotes the
    finger growth for glycerol while dashed line denotes the finger
    growth for PDMS.}
\end{figure}

Finally, we discuss the growth of the fingers post-instability
formation. We consider only the fastest growing finger, though the
behavior is more or less the same for other
fingers. Figure~\ref{fig:front_position} shows the growth of the
fastest growing finger ($R_{f} - R_{c}$) normalized by the cube root
of the initial volume ($V_{0}$) at $500$ rpm. Here, $R_{f}$ is the
distance of the finger front from the axis of rotation and $R_{c}$ is
the critical radius. Glycerol forms less number of fingers and all the
fluid passes outwards from the existing fingers leading to a faster
growth rate (solid line in Fig.~\ref{fig:front_position}). In
contrast, PDMS forms more fingers and the liquid flows outward through
fingers as well as during the slow spreading of the central drop
causing an overall lower rate in the finger growth (dashed line in
Fig.~\ref{fig:front_position}). With increasing particle fractions,
the suspension data show a monotonic decrease in the rate at which
the fingers grow in time. The rate, is faster at lower values of
$\phi_{p}$, decreases and is nearly identical to that for glycerol at
$\phi_{p}=0.475$, decreases further to be similar to that for PDMS at
$\phi_{p} = 0.5$ and falls further for highest values of $\phi_{p}$
studied. This overall behavior is a combination of varying viscosity
and the particle influence on finger growth with increasing
$\phi_{p}$.

\section{Conclusions}
\label{sec:conclusions}

In summary, we have shown, through flow visualization experiments, a
unique and curious effect of the particle fraction on the spreading
behavior of a film of suspension. The critical radius for the onset of
instability is found to increase gradually with increase in the
particle fraction with a slight drop off at the highest particle
fraction studied, while the instability wavelength decreases and then
increases again while going through a minimum for the same range of
particle fractions. The decrease and increase in the wavelength are,
respectively, attributed to the possible particle concentration
variations in the capillary ridge at lower $\phi_{p}$ and particle
crowding at higher $\phi_{p}$. The supporting arguments provided on
the basis of previous theoretical
studies~\cite{Balmforth2007,Cook2009} are qualitative, but the results
presented are, nevertheless, quite significant. They should provide
strong impetus towards developing a theoretical framework capable of
encompassing the existence of seemingly different spreading
characteristics of a suspension film within a small concentration
range. Some interesting avenues within the scope of future work are
(i) actual visualization of particle motion within the capillary ridge,
(ii) changes in the contact line instability for a mixture of particles
differing in size, for non-neutrally buoyant particles and for
non-Newtonian suspending liquids, and (iii) altering the particle surface
properties to change its affinity with respect to liquid. The
results also suggest possible mechanism of altering the spreading
behavior of a thin film of liquid by addition of non-interacting
particles which should be of interest to industrial applications.
 
\begin{acknowledgments} We thank Ashish Lele, Ganesh Subramaniam,
  Mahesh Tirumkudulu, Prabhakar Ranganathan, Prabhu Nott, and Sarika
  Bhattacharya for several fruitful discussions, Arun Banpurkar for
  providing help in measuring the surface tension of all the fluids,
  and Sameer Huprikar for the help provided in rheology
  measurements. The financial support from the Department of Science and
  Technology, India (Grant No. SR/S3/CE/0044/2010) is gratefully
  acknowledged. \end{acknowledgments}

\bibliography{paper.bib} 

\end{document}